\documentstyle[aps,multicol,epsf]{revtex}
\input epsf
\draft
\title{Domain Walls Motion and Resistivity in a Fully-Frustrated  
Josephson Array}
\author{M.V. Simkin\footnote{email: simkin@physics-sun.rockefeller.edu}\\ 
{\em Department of Physics, Brown University, Providence, RI 02912-1843}\\
{\em The Rockefeller University, New York, NY 10021-6399}\footnote{present
address}}
\date{}
\begin{document}
\maketitle
\begin{abstract}
It is  identified numerically that the resistivity of a fully-frustrated 
Josephson-junction array is due to 
motion of domain walls in vortex lattice rather than to motion of single 
vortices.
\end{abstract}
\begin{multicols}{2}
\narrowtext

A square two-dimensional periodic Josephson-junction array (JJA) in a 
uniform
transverse  magnetic field, with half a flux quantum per plaquette, is a 
realization of a fully frustrated (FF) $XY$ model \cite{vil}.
Its ground state is a checkerboard pattern of plaquettes with currents
flowing clockwise and anti-clockwise \cite{tei}.The chirality, defined 
as the sum of currents around plaquette in a clockwise direction, has 
antiferromagnetic order. Apart from Kosterlitz-Thouless (KT) \cite{kt} 
transition 
as in the unfrustrated case the FF $XY$ model has an Ising-like transition 
of 
chiral ordering \cite{tei2},\cite{ber}.  Numerical 
\cite{tei2}~-~\cite{olson} and
analytical \cite{kg} work indicates that critical temperatures of KT and 
Ising
transitions are very close or equal.
Mon and Teitel \cite{mt} studied current-voltage curves of the FF
JJA using Langevin-dynamics simulation. They found that the temperature 
dependence of the exponent $a$ of
the current voltage characteristics ($V \sim I^{a+1}$) does not follow the 
form
expected from the results of equilibrium simulations of helicity modulus 
of FF$XY$ model
\cite{tei2}. In particular there were no sign of discontinuity in $a$ at 
the critical temperature, 
while the  discontinuity  was observed in the helicity modulus 
\cite{tei2}. Mon and Teitel \cite{mt} proposed that this discrepancy
is because apart from vortex-antivortex pairs (KT excitations)
the system has excitations in the form of domain walls in vortex lattice 
(Ising excitations), motion of which  also contributes to resistivity. 
Disagreement between  the exponent of the current voltage characteristics 
of FF JJA and helicity modulus of  FF$XY$ model has been also observed 
experimentally, when in some cases \cite{vdz0} no sign of discontinuity 
in $a$ was seen , and in others the sharp jump in $a$ was observed 
\cite{vdz2}, but it was substantially below $T_c$, expected from simulations.
In contrast for the unfrustrated case \cite{vdz1} the sharp jump in $a$ 
was observed at the temperature equal to $T_c$, found in  simulations. 
Some experiments\cite{ling} on superconducting wire networks in a 
magnetic field have reported exponential, instead of power-law, $I-V$ 
curves. The exponential form is expected theoretically \cite{mt} if 
resistivity is due to motion of thermally activated domain-walls in vortex 
lattice.

In this article it is  shown, by numerical simulations, that domain-wall
motion dominates the resistivity at least at low temperatures and small 
currents.

In simulations the  modification of  Langevin dynamics
method of Refs. \cite{mt} and \cite{str} proposed by Falo, Bishop, and Lomdahl
\cite{fal} was used.
All junctions are assumed to have the same critical currents $I_0$ and 
to be shunted by resistances $R$ and all
superconducting nodes to have a capacitance to the ground $C$.
The resulting set of dynamic equations (which follows from Josephson
equations and the charge conservation law) is:

\begin{equation}
	\frac{d\theta_{{\mathbf n}}}{dt}=\frac{2e}{\hbar}P_{{\mathbf n}},
\end{equation}

\begin{eqnarray}
	\frac{dq_{{\mathbf n}}}{dt}=I_0\sum_{<{\mathbf m}>}
	\sin(\theta_{\mathbf m}-
	\theta_{\mathbf n}- A_{\mathbf mn})+ \nonumber \\
\frac{1}{R}\sum_{<{\mathbf m}>}(P_{{\mathbf m}}-
	P_{{\mathbf n}})+ \sum_{<{\mathbf m}>}I^{fl}_{{\mathbf mn}},
\end{eqnarray}

\begin{equation}
	P_{{\mathbf n}}=\frac{q_{\mathbf n}}{C}.
\end{equation}
Here $\theta_{\mathbf n}$ is the phase, $P_{{\mathbf n}}$ is the 
electrostatic potential
and $q_{{\mathbf n}}$ is the charge of the superconducting node on site 
${\mathbf n}$,
$A_{\mathbf mn}=\frac{2e}{\hbar c}\int_n^m {\mathbf A}d {\mathbf l}$ is the 
integral
of the vector potential from node $n$ to $m$ (sum of $A_{\mathbf mn}$ around 
the
plaquette is equal to the number of magnetic flux quantums through the 
plaquette).
Summation is over nearest neighbors only. $I^{fl}_{\mathbf mn}$ is a thermal 
noise current with:
\begin{equation}
	<I^{fl}_{\mathbf mn}(t)I^{fl}_{\mathbf mn}(\tilde{t})>=
	(2T/R)\delta(t-\tilde{t}),
\end{equation}
with $T$ the temperature.

\begin{figure}
\centering	
\begin{minipage}{7cm}
\epsfxsize= 7cm \epsfbox{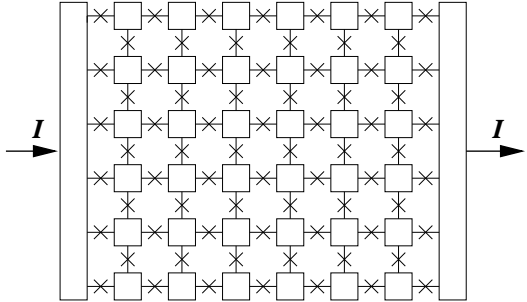}
\end{minipage}
\vspace{1cm}
\caption{Josephson-junction array with superconducting bars at the edges.
The external current is injected into the left bar and extracted from the 
right. Voltage is measured between the bars.}
\label{array}
\end{figure}

Instead of the uniform injection of Ref. \cite{mt} injection from
the superconducting bars at the edges of the array (see Fig. \ref{array}) was 
used in 
this work in order to imitate the experiment \cite{vdz2}, \cite{vdz1}.
 Free boundary conditions in the transverse
direction where used for the same reason.
The bars were connected to each of the nodes at the edges of the array
by Josephson junctions. The external current was injected into the left bar
and extracted from the right bar. It was observed \cite{free} that for the
FF JJA injection 
from bars leads to unreasonably low zero-temperature critical current in 
comparison to uniform injection. The reason for this is that FF plaquettes
adjacent to bars have only three Josephson-junctions leading to bigger phase
differences in the ground state and therefore lower critical currents.
In this work plaquettes adjacent to bars were assumed unfrustrated 
(experimentally this would correspond to this plaquettes having an area much
smaller than plaquettes in the bulk) and the zero-temperature critical current
was the same as in the case of the uniform injection ($I_c \cong 0.35$), 
though lower than the theoretical value for an infinite array ($I_c =
\sqrt{2}-1$)\cite{bnz},\cite{hal}.

Simulations were done for the intermediate damping case, i.e. for McCumber
parameter $\beta=1$. The equations of motion were integrated with discrete
time steps $\Delta t =0.05$ in units of inverse Josephson plasma frequency
$\omega_J=\sqrt{2eI_0/\hbar C}$. Decreasing  $\Delta t$  ten times did not 
change the results.

\begin{figure}
\centering	
\begin{minipage}{7cm}
\epsfxsize= 7cm \epsfbox{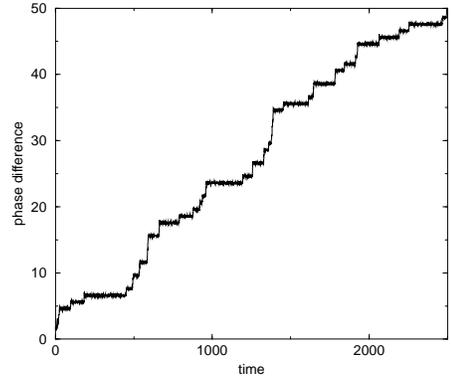}
\end{minipage}
\begin{minipage}{7cm}
\epsfxsize= 7cm \epsfbox{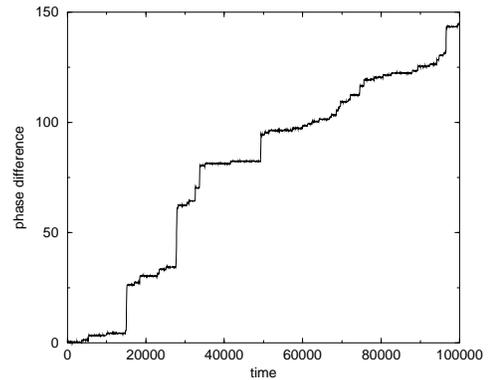}
\end{minipage}
\caption{(a)Random staircase is the phase difference (in units of $2\pi$)
across the $16 \times 16$ 
unfrustrated array versus time (in units of inverse plasma frequency) 
for temperature $T=0.3$ and current $I/I_{0}=0.5$. It is clearly seen that
the phase difference grows by $2\pi$ jumps, corresponding to vortex
crossings shown in  Fig.4. (b)The same for the $17 \times 17$ 
fully-frustrated array  for temperature $T=0.3$
and current $I/I_{0}=0.07$. One can see the giant ($ \sim 20 \times 2\pi$)
jumps corresponding to domain wall propagation shown in Fig.5.}
\label{staircase}
\end{figure}

The phase difference $\phi$ across the array was recorded as a function of 
time (see Fig. \ref{staircase}). The average voltage is:
\begin{equation}
	\frac{V}{RI_0}=\frac{\phi_{end}-\phi_{begin}}{t_{run}},
\end{equation}
where $\phi_{begin}$ and $\phi_{end}$ are the phase differences across the 
array at the beginning and the end of the run correspondingly, and $t_{run}$
is the run time (in units of $1/\omega_J$). 
To estimate the errors 
each run was divided in four intervals
and  average voltages for this intervals were calculated. The estimated errors
are  $\frac{\Delta V}{V} \sim 0.1$.

In the following we shall adopt the following units:
the current is measured in units of the critical current of the unfrustrated
array $I_0$, the voltage in units of $RI_0$, and the temperature 
in units of the Josephson energy $\hbar I_0/2e$.

\begin{figure}
\centering	
\begin{minipage}{7cm}
\epsfxsize= 7cm \epsfbox{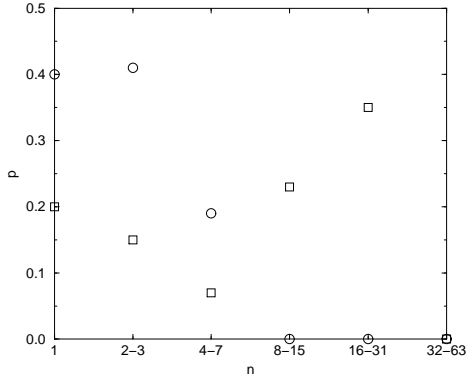}
\end{minipage}
\caption{A fraction, $p$, of the total phase slip in Fig. 2,
due to jumps of size $n2\pi$. Unfrustrated array - circles, 
fully-frustrated array - squares.}
\label{histogram}
\end{figure}

Looking at Figure \ref{staircase} (a) and (b), where phase difference across 
unfrustrated and
frustrated arrays is plotted versus time, one can notice a remarkable 
difference. While the phase difference across the unfrustrated array grows by
$2\pi$ jumps, the one across the fully-frustrated array grows by giant
$\sim 20 \times 2\pi$ jumps. In order to quantify results more we classify
jumps according to their size.
A jump of size $2\pi n$ is put into bin $k$ such that 
$2^{k-1} \leq n < 2^k$. A fraction of the total phase slip, due to
jumps in each bin is shown in Fig. \ref{histogram}. We see that for
unfrustrated array jumps in 1st and 2nd bins (that is $n=1,2,3$) dominate,
while for the fully-frustrated array jumps in 5th bin ($n=16, \ldots, 31$)
give biggest contribution to the total phase slip.

As the Langevin dynamics simulation is performed
using {\it pseudo}-random numbers one can always go back and investigate what
happens in the vicinity of the jump in detail. In Fig. 4 the 3 snapshots of
the unfrustrated array in the region of one of the jumps are shown. One can 
clearly see a vortex crossing the array. Therefore the resistivity in the
unfrustrated array is due to vortex crossing \cite{com}. 
In Fig. 5 the 3 snapshots of the FF array in the region of the giant jump
are shown. It is hard to see what is going on in the FF array in the phase
representation, therefore the chirality (sum of currents around plaquette in
a clockwise direction) representation was used.
However even then domain walls are not readily seen. To make them clearly
visible one ground state is represented by vertical and other by horizontal
arrows. One can see that before the giant jump the array is in one of chiral 
ground states, during the jump the domain of opposite chirality grows and
at the end of the jump it fills all of the array. Resistivity of the FF array
is therefore due to domain wall motion.

The method described here works, however, only for low temperatures 
($T < 0.35$)and low currents ($I < 0.1$). When the temperature or current  is 
increased one has difficulties 
in identifying the separate jumps, while the phase difference - time curve
is smooth.

\begin{figure}
%\centering	
%\begin{minipage}{16.5cm}
\epsfxsize= 5.cm 
\epsfbox{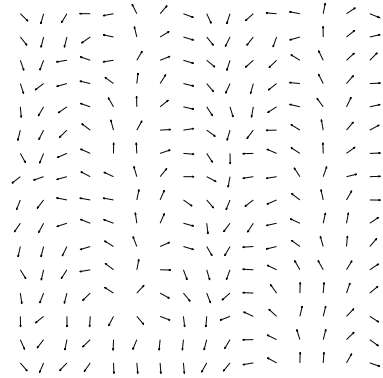}
\epsfxsize= 5.cm 
\epsfbox{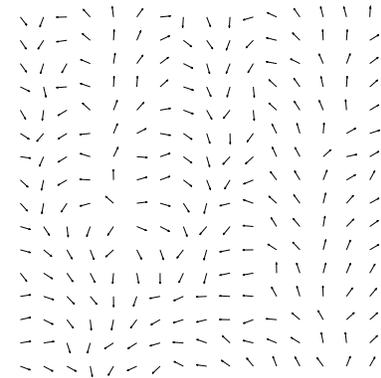}
\epsfxsize= 5.cm 
\epsfbox{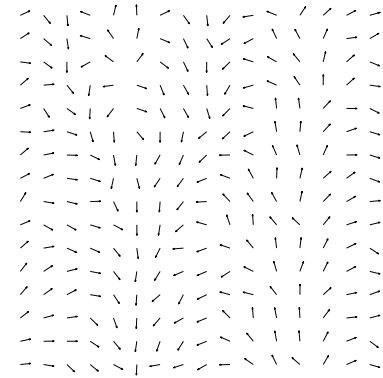}
%\end{minipage}
\caption{Vortex crossing in the unfrustrated array of size $16 \times 16$. 
Phases of superconducting
grains are indicated. Three consecutive snapshots (a),(b),(c) are in the 
region of the phase jump at time $t \sim 100$ (see Fig.2(a)).}
\end{figure}

\begin{figure}
%\centering	
%\begin{minipage}{18cm}
\epsfxsize= 5.cm \epsfbox{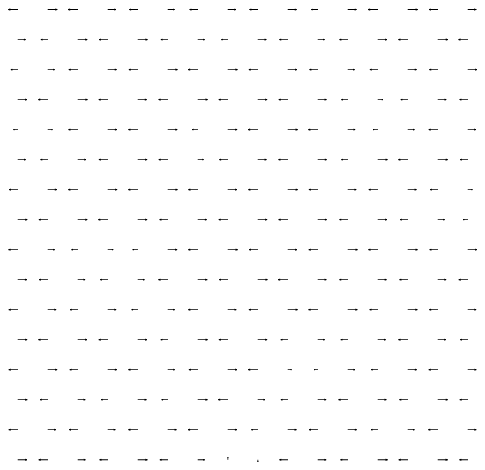}
\epsfxsize= 5.cm \epsfbox{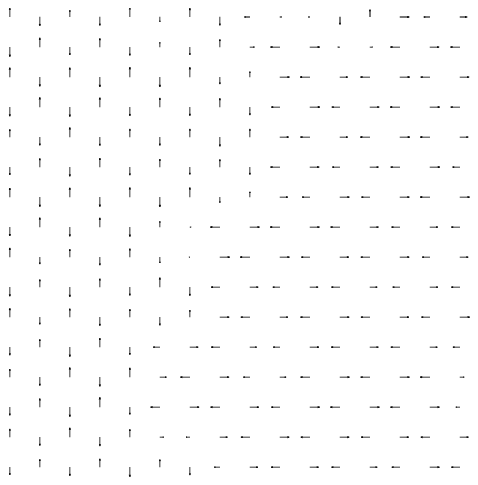}
\epsfxsize= 5.cm \epsfbox{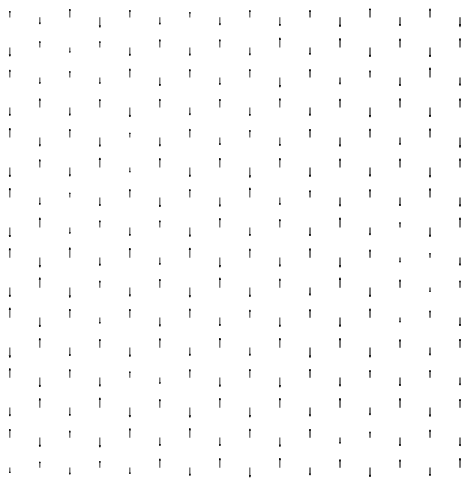}
%\end{minipage}
\caption{Domain wall propagation in the fully-frustrated array of size 
$17 \times 17$.
Chirality is indicated. Two ground states are represented by vertical and
horizontal arrows. Three consecutive snapshots (a),(b),(c) are in the 
region of the giant phase jump at time $t \sim 1.5 \times 10^4$ 
(see Fig.2(b)).}
\end{figure}

\begin{figure}
\centering
\begin{minipage}{7.0cm}
	\epsfxsize= 7cm \epsfbox{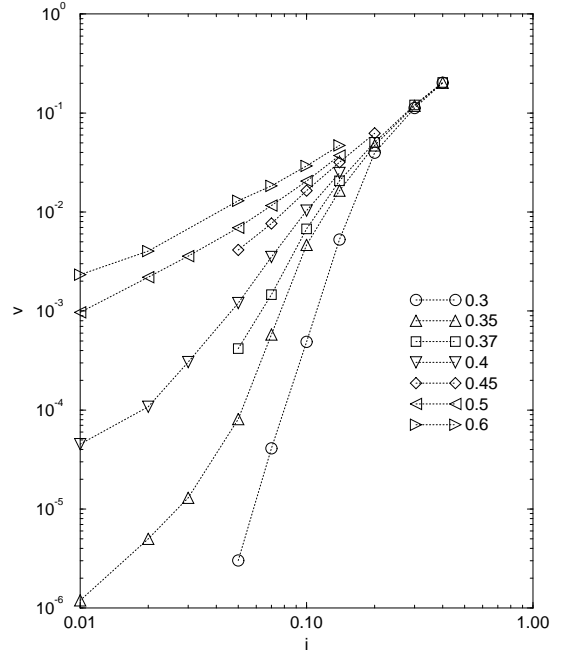}
\end{minipage}
\caption{{\it I-V } curves ($i=I/I_0$ vs $v=V/RI_0$) of the fully-frustrated 
Josephson-junction array for 5 different temperatures (given in units of
Josephson energy, $\hbar I_0/2e$).}
\end{figure}

\begin{figure}
\centering	
\begin{minipage}{8cm}
\epsfxsize= 8cm \epsfbox{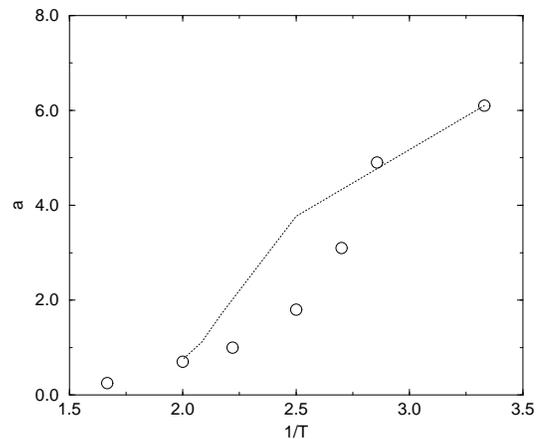}
\end{minipage}
\caption{Exponent $a$ of the current-voltage characteristics 
($V \sim I^{a+1}$)
versus inverse temperature ($1/T$), obtained from the data of Fig.6. 
Dashed line is obtained from helicity modulus (Ref. [4]) using Eq. 6.}
\end{figure}

The $I-V$ curves of the FF array for different temperatures are shown in 
Fig. 6. And the exponent $a$ ($V \sim I^{a+1}$), obtained from them, is shown 
in Fig. 7. There is a good agreement between the data of Fig.6 and those
of Mon and Teitel (Fig.1(b) of Ref. \cite{mt}).  There is less agreement
between the exponent $a$ obtained in this work (Fig. 6) and that of 
Mon and Teitel (Fig. 2 of Ref. \cite{mt}). This is due to the fact that much 
lower currents have been studied in the present work than in Ref. \cite{mt},
where  $a$ was obtained by fitting the $I-V$ curves in the region of current
just below the critical one.
In this work $a$ was obtained by fitting in the region of current 0.05-0.1
(in units of the critical current of the unfrustrated array).  For higher
current single-junction effects are important, for lower ones - finite
size effects \cite{ks} cause the change of the $I-V$ curve from nonlinear to 
linear. 
It is interesting, that though the resistivity is due to domain walls motion
the $I-V$ curves  still can  be fitted by the power law. 
In Fig. 7 the theoretically expected \cite{hn} exponent of the $I-V$ relation,
calculated as 
\begin{equation}
	a=\pi \Gamma /T,
\end{equation}
 using helicity modulus $\Gamma$ obtained in 
Ref. \cite{tei2} is shown by dashed line. Though agreement between  $a$ 
predicted using helicity modulus and obtained from $I-V$ curves of the present
work is better than in Ref. \cite{mt}, still there is disagreement. The reason
for it is probably domain wall contribution to resistivity, as suggested
by Mon and Teitel \cite{mt} and demonstrated explicitly in the present work.

In conclusion in this work an approach is suggested when one records phase
difference across the Josephson array versus time (time derivative of the
phase difference is the voltage).
Resulting pictures are
different in the cases of unfrustrated and fully frustrated arrays. In the 
first case phase difference grows by $2\pi$ jumps, while in the last by
giant (many $2\pi$) jumps. Snapshots of the array in the region of a giant jump
show that during it there is a transition from one of two chiral ground state
to another, therefore identifying the mechanism of resistivity as domain
walls motion.

I am grateful to J.M. Kosterlitz  for useful conversations and to E.Granato
and S.Teitel for correspondence. This work was supported by 
NSF Grant No. DMR-9222812 and DOE Contract No. DE-F602-88-ER13847. 
Computations  were done at the Theoretical Physics Computing Facility at 
Brown University.

\end{multicols}

\begin{thebibliography}{99}
\bibitem{com} Larger than $2 \pi$ jumps in the unfrustrated
array may be due to the fact that pair unbinding  occurs
when temperature rises due to a fluctuation, however in this case other
pairs in the array  are likely to unbind as well.
\end{thebibliography}
\end{document}